\RequirePackage{booktabs}

\documentclass[sn-mathphys,Numbered,iicol]{sn-jnl}

\usepackage{array}
\usepackage{enumitem}
\usepackage[ruled,linesnumbered,algo2e]{algorithm2e}

\usepackage{lineno} 

\usepackage{graphicx}%
\usepackage{multirow}%
\usepackage{amsmath,amssymb,amsfonts}%
\usepackage{amsthm}%
\usepackage{mathrsfs}%
\usepackage[title]{appendix}%
\usepackage{xcolor}%
\usepackage{textcomp}%
\usepackage{manyfoot}%
\usepackage{booktabs}%
\usepackage{algorithm}%
\usepackage{algorithmicx}%
\usepackage{algpseudocode}%
\usepackage{listings}%

\definecolor{mygreen}{rgb}{0,0,0}

\theoremstyle{thmstyleone}%
\theoremstyle{thmstyletwo}%

\theoremstyle{thmstylethree}%
\newtheorem{definition}{Definition}%

\begin{document}

\title[Article Title]{MEGG: Replay via Maximally Extreme GGscore in Incremental Learning for Neural Recommendation Models}


\author[1]{ \sur{Yunxiao~Shi}}\email{Yunxiao.Shi@student.uts.edu.au}

\author[1]{ \sur{Shuo~Yang}}

\author[1]{\sur{Haimin~Zhang}}

\author[1]{\sur{Li~Wang}}

\author[1]{\sur{Yongze~Wang}}

\author[1]{\sur{Qiang~Wu}}

\author*[1]{\sur{Min~Xu}}\email{min.xu@uts.edu.au}

\affil*[1]{\orgdiv{School of Electrical and Data Engineering}, \orgname{University of Technology Sydney}, \orgaddress{\street{Broadway}, \city{Sydney}, \postcode{2007}, \state{NSW}, \country{Australia}}}


\abstract{
\textcolor{mygreen}{Neural Collaborative Filtering (NCF)-based recommendation models have been widely adopted in practical recommender systems due to their effectiveness. However, these models are typically developed under the static deep learning paradigm, where training is conducted on fixed datasets with the implicit assumption of a static data distribution. This approach is ill-suited for dynamic environments, such as those encountered in real-world platforms, where user preferences and collaborative filtering patterns evolve continuously. To address this limitation, incremental learning—a paradigm designed to integrate new knowledge while preserving previously learned information—emerges as a promising alternative. Despite its potential, the direct application of conventional incremental learning methods, which are prevalent in domains like computer vision and natural language processing, is hindered by unique challenges in recommender systems. These include the distinct task paradigm, data complexity, and sparsity issues. Moreover, existing incremental learning approaches tailored for neural recommendation models remain scarce and often suffer from limited generalizability. To bridge this gap, we propose an innovative experience replay-based incremental learning framework specifically designed for neural recommendation models, termed Replay Samples with Maximally Extreme GGscore (MEGG). At the core of MEGG is a novel metric, the GGscore, which quantifies the influence of individual samples on model training. By selectively replaying samples with the most extreme GGscores, our method effectively mitigates catastrophic forgetting, thereby maintaining high predictive performance over time. A key advantage of MEGG lies in its data-centric nature, which renders it agnostic to the underlying model architecture. This ensures broad applicability across various neural recommendation models and seamless integration with existing incremental learning frameworks to further enhance performance. Extensive experiments conducted on three neural recommendation models across four benchmark datasets demonstrate the superior effectiveness of MEGG compared to state-of-the-art methods. Furthermore, additional evaluations highlight its scalability, efficiency, and robustness. The implementation of MEGG will be made publicly available upon acceptance.}
}




\maketitle

\clearpage
\section{Introduction}\label{sec1}
\textcolor{mygreen}{
Recommender systems are widely used across a broad range of applications, with recommendation algorithms serving as their core.  Among the myriad of algorithmic paradigms, recommendation models based on deep neural networks (commonly referred to as Neural Collaborative Filtering, or NCF \cite{NCF}) have garnered significant traction within the industry due to their implementation simplicity and high efficiency in delivering effective results \cite{wide_deep,Deep_crossing,neural_FM,DeepFM,xDeepFM,DIN,dlrm,NCF}. Traditionally, these recommendation algorithms follow the conventional deep learning paradigm, where models are trained on fixed datasets and then applied to unseen data under the assumption of a static data distribution. However, in many real-world applications, such as music streaming \cite{lee2007context}, news recommendation \cite{liu2010personalized}, Point-Of-Interest (POI) recommendation \cite{chang2020learning}, movie recommendation \cite{movie_rec_evolve}, and e-commerce platforms \cite{sarwar2000analysis}, recommender systems operate in dynamic environments where user interaction data stream is continuously generated \cite{ADER, SML, FIRE}, reflecting the evolving nature of users' preferences. This implies that incoming streaming data, which has not been observed during training, may differ significantly from the original training data in terms of distribution. As a result, models previously trained in static environments, when deployed under dynamic conditions for extended periods, often experience a decline in predictive performance \cite{graphsail}.
}


The paradigm of incremental learning \cite{IL_ECCV2018, Three_scenarios_CL, class_IL_survey1, class_IL_survey2, IL_OD}, which entails the continual acquisition of new knowledge while preserving previously learned information, offers a compelling framework for the development of dynamically evolving recommender systems. \textcolor{mygreen}{Although incremental learning has been extensively investigated in domains such as natural language processing \cite{CL-NLP, jang2022towards} and computer vision \cite{CL_CV_review, mai2022online, wang2022continual, class_IL_survey2}, its adoption to neural recommendation models is still in its formative stages. A primary obstacle lies in the fundamental differences in task structures. Mainstream incremental learning methods are predominantly designed to address Class-Incremental and Task-Incremental scenarios \cite{Three_scenarios_CL}. In contrast, recommender systems operate within a distinct Domain-Incremental paradigm characterized by a continuous influx of streaming bipartite user-item interaction data. This data reflects evolving patterns of collaborative filtering, driven by the dynamic and perpetual shifts in user preferences. Furthermore, data in the recommendation domain is inherently heterogeneous and sparse, with each interaction often encapsulating complex, multifaceted relationships among multiple entities, in stark contrast to the typically isolated data in domains like computer vision and natural language processing. These complexities underscore highlights the imperative for designing novel methodologies capable of accommodating the incremental and adaptive demands of recommendation tasks.}

\begin{figure*}[t]
  \centering
    \includegraphics[width=5.0in]{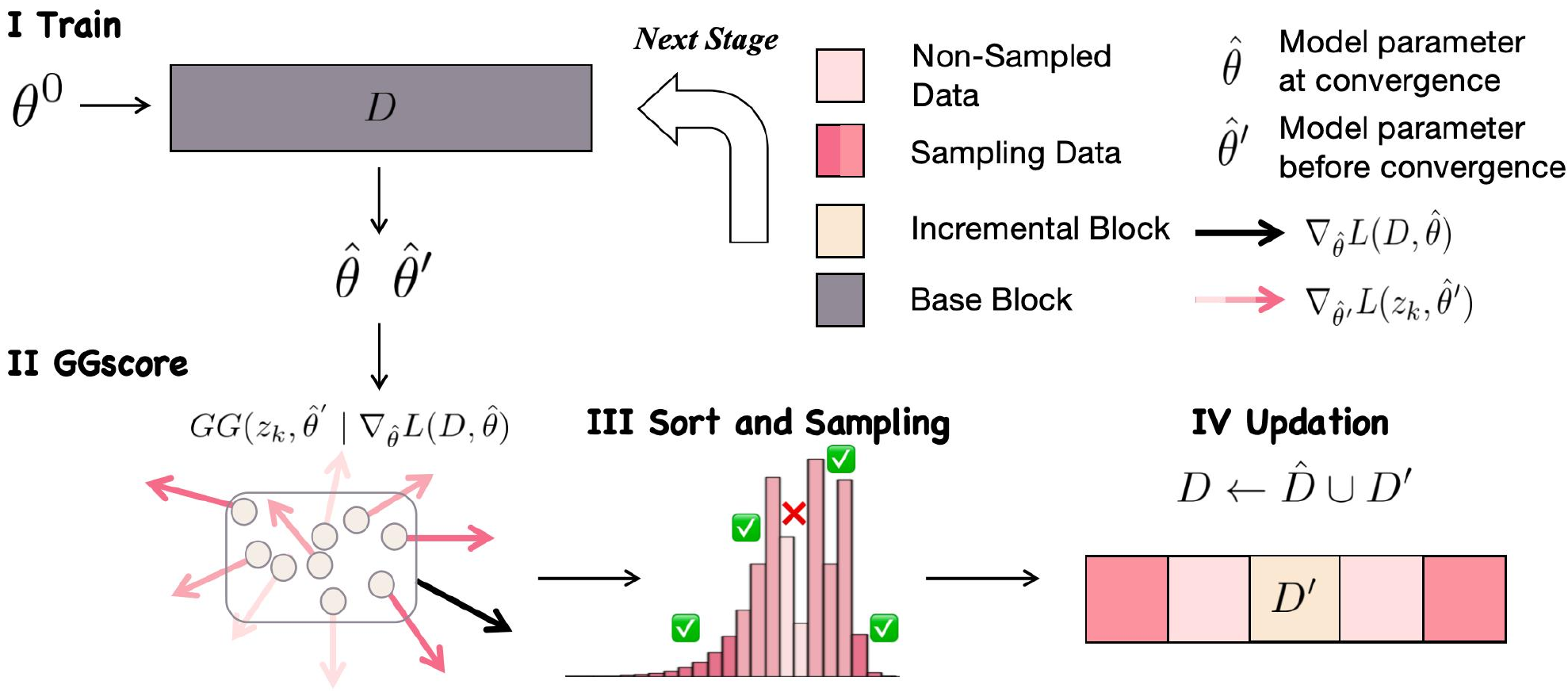}
    \vspace{-2mm}
    \textcolor{mygreen}{
   \caption{Overview of the MEGG framework for incrementally training neural recommendation models. The framework addresses dynamic environments where streaming data continuously arrives at each stage as the incremental block \(D'\). At the start of each stage, model parameters are initialized as \(\theta^0\) and trained on $D'$ to convergence, yielding \(\hat{\theta}\), and an intermediate set of parameters \(\hat{\theta}'\) is also stored for computing the GGscore, which quantifies the impact of specific user-item interactions \(z_k\). We prioritize sampling influential records for replay.}
   }
    \label{fig:framework}
\end{figure*}

\textcolor{mygreen}{Therefore, there has been an increasing emphasis on the development of incremental learning techniques tailored specifically for recommender systems. Notable advancements include approaches designed for graph-based recommendation models \cite{graphsail,graph_structure_aware_replay,Graph_Contrastive_Knowledge_Distillation,FIRE,casual_IGCN}, as well as methods targeting generative recommendation paradigms \cite{ACN,ERAGent,zhang2023memory,hatalis2023memory}.  However, comparatively fewer efforts have been directed toward neural recommendation models \cite{taboola_IRS,huawei_IRS,linkedin_IRS,IncFM,fast_inc_FM}. Taboola introduced an incremental learning framework for Click-Through Rate (CTR) prediction called CTR-IL \cite{taboola_IRS}, which leverages knowledge distillation to transfer knowledge from the old model to the new one. This approach is advantageous due to its ease of implementation and practical applicability.  However, student model often fail to achieve the performance levels of the teacher model \cite{KD_survey}, leading to suboptimal recommendation performance. Similar to CTR-IL, Huawei also proposed a method called IncCTR \cite{huawei_IRS}, in addition to applying knowledge distillation technique, it conceptually introduces a data module designed for replaying samples, for further improving the performance. But this module is only conceptual and has not been deeply investigated. Moreover, LinkedIn \cite{linkedin_IRS} approaches incremental learning as a Sequential Bayesian Update problem, thereby enabling logistic regression models to adapt incrementally. Nevertheless, the foundational assumptions of this method may not be fully applicable to neural network models. As a result, the generalizability of this approach is somewhat constrained, limiting its broader applicability within the context of neural recommendation models.}

\textcolor{mygreen}{
To advance the exploration of this critical domain, we propose a novel and orthogonal approach to existing methods. Specifically, our work adopts a data-centric viewpoint, focusing on the experience replay paradigm \cite{class_IL_survey1,class_IL_survey2}, a category of methods that emphasizes the effective utilization of historical data. Unlike model-dependent strategies, data-centric approaches are inherently agnostic to the underlying architecture, offering broad applicability across diverse neural recommendation models. Our approach is also a practical implementation of the conceptual data module introduced in IncCTR \cite{huawei_IRS}, the modular can be seamless integrated into existing neural recommendation systems, ensuring ease of implementation and operational scalability. We start by implementing the conceptual data module using several well-established experience replay methods from the general incremental learning domain, such as Gdumb \cite{Gdumb}, iCaRL \cite{iCaRL}, and MIR \cite{MIR}. Building on these foundations, we proceed to develop a novel sampling strategy tailored specifically for incremental learning in neural recommendation models.} As a cornerstone of this strategy, we introduce a new concept named \textit{One Step Loss Change}, which is designed to quantifies the change in training loss during one optimization step caused by the removal of a single user-item interaction sample. Through rigorous empirical investigation, we have discovered its efficacy as a robust criterion capable of assessing the influence of samples on model training. Based on this observation, our research aims to address the following question: \textbf{How can we leverage the property of One Step Loss Change to select important user-item records for experience replay?} We outline our approach in four steps to address this problem. Firstly in Section~\ref{sec:theo}, we estimate One Step Loss Change in Mini-Batch Gradient Descent settings and find that the estimation formula constituted of the dot product between several gradients. Subsequently in Section~\ref{sec:GGscore}, to further simplify the measurement of influence, we propose a criterion called GGscore $GG(z,\theta \mid \mathbf{V})$, which quantifies the alignment between the gradients of a sample $z$ with respect to the given model parameters $\theta$ and a given reference vector $\mathbf{V}$ of the same dimensionality. Moreover, in Section~\ref{sec:parameter_selection}, to improve the computing efficiency of GGscore and make it applicable to large-scale recommendation models, we propose to compute the gradient only on partial parameters of models. Finally in Section~\ref{sec:replay_MEGG}, we propose an experience replay-based incremental learning method that retains samples with Maximally Extreme GGscore (MEGG) for future periodic replay. We present the framework illustrating the application of MEGG in a dynamic recommender system in Figure~\ref{fig:framework}. The results of extensive experiments conducted on three neural recommendation models across four different-size datasets demonstrate the following characteristics of the MEGG method: effectiveness (in Section~\ref{sec: performance_comparison} and Section~\ref{sec: integration}), expandability (in Section~\ref{sec: integration}), efficiency (in Section~\ref{sec: efficiency_analysis}), and robustness (in Section~\ref{sec: hyperparameter_analysis} and Section~\ref{sec: sampling_size_analysis}).

We summarize our contributions as threefold:

(1) We propose a novel criterion named GGscore to quantify the influence of individual samples on model training. Additionally, leveraging the parameter-affected locality property of neural recommendation models, we design the GGscore to be in low-parameters computational requirement, ensuring its high-velocity in large-scale recommender systems.

(2) We have contributed to the advancement of the field of incremental learning for recommender systems by introducing an orthogonal approach called Replay via Maximally Extreme GGscore (MEGG). Notably, MEGG is the pioneering experience replay-based incremental learning method specifically tailored for neural recommendation models.

(3) Extensive experiments confirm that MEGG is a very potential method to address catastrophic forgetting in recommender systems with good effectiveness, expandability, efficiency, and robustness.

\section{Related Work}
\subsection{Incremental Learning}
Incremental Learning (IL), also referred to as Continual Learning (CL) \cite{IL_ECCV2018,catastrophic_forgetting,ICML2017_Continual_Learning,NIPS2017_Continual_Learning,Three_scenarios_CL}, enables artificial intelligence models to continuously acquire and integrate new knowledge over time. A central challenge in IL is addressing the issue of catastrophic forgetting \cite{catastrophic_forgetting}, wherein models, while adapting to new information, lose previously acquired knowledge, leading to a significant degradation in performance on earlier tasks. The IL paradigm is typically categorized into three scenarios \cite{Three_scenarios_CL}: Task-Incremental Learning (Task-IL), Class-Incremental Learning (Class-IL), and Domain-Incremental Learning (Domain-IL). \textcolor{mygreen}{Initially, IL research gained traction in the field of computer vision (CV) \cite{CL_CV_review,mai2022online,wang2022continual,class_IL_survey2}, and then its applications have expanded into natural language processing (NLP) \cite{CL-NLP,jang2022towards}, with most studies focusing on Task-IL and Class-IL scenarios.} However, in recommendation systems, the challenges of IL align more closely with the Domain-IL scenario. Despite the growing importance of incremental learning in recommendation contexts, research efforts tailored specifically to this domain remain sparse. 
Notably, recent advances have explored adapting incremental learning (IL) methodologies to graph-based recommendation models \cite{graphsail,graph_structure_aware_replay,Graph_Contrastive_Knowledge_Distillation,FIRE,casual_IGCN}, \textcolor{mygreen}{as well as incorporating memory modules \cite{ACN,ERAGent,zhang2023memory,hatalis2023memory} to enable generative recommendation based on large language models \cite{llmrank,llmrec,SLMRec} to incrementally capture users' evolving preferences. However, due to the inherent discrepancies in paradigms, the aforementioned incremental learning approaches cannot be seamlessly adapted to neural recommendation models. Research specifically addressing this point remains in its infancy, with only a handful of pioneering studies \cite{huawei_IRS,taboola_IRS,linkedin_IRS,SML} venturing into this domain. }

IL methodologies can be classified into three principal approaches \cite{kemker2018measuring, class_IL_survey1, ICML2017_Continual_Learning, classInc_comparative_survey}: regularization-based methods, experience replay strategies, and model isolation techniques. Among these, experience replay method is rooted in the data-centric AI principles \cite{data_centric}, uniquely focuses on reusing past data. Studies have shown that hybrid approaches integrating experience replay mechanism consistently achieve improved performance \cite{classInc_comparative_survey, daxberger2023improving}. \textcolor{mygreen}{In this study, we explore an incremental learning method for neural recommendation models from the perspective of experience replay. The aim is to develop a method that is widely applicable, whether for any type of neural model architecture or for existing incremental recommendation frameworks, enhancing or further improving their performance. Specifically, we first attempt to adapt several established experience replay methods from general incremental learning, including Gdumb \cite{Gdumb}, iCaRL \cite{iCaRL}, and MIR \cite{MIR}, to recommendation scenarios. Building upon this foundation, we introduce MEGG, pushing the boundaries of current state-of-the-art techniques in this domain.}

\subsection{Sampling Strategy in Experience Replay}
\label{sec: incremental_learning}
Experience replay-based incremental learning methods mitigate catastrophic forgetting by retaining a subset of historical data and periodically replaying it while learning new information. Currently, the sampling strategy can be bifurcated into two divergent perspectives. Represented by iCaRL \cite{iCaRL}, which emphasizes the retention of highly prototypical instances. Conversely, the alternative viewpoint presents in MIR \cite{MIR} advocates preserving the samples that lie close to the classification decision boundaries. But Gdumb \cite{random_replay_effective, Gdumb} indicates that the performance of these prevailing methods is even inferior to that of a simple random sampling strategy. \textcolor{mygreen}{In our study, we are not concerned with whether a sample is prototypical or lies near the decision boundary. Instead, we focus on its influence—if removing a sample has no effect on the model's predictive performance, we consider that sample redundant. The remaining samples, by contrast, are deemed more important and should be retained. Our method is more intuitive and explainable.}

\section{Preliminaries}

\subsection{Experience Replay Procedure}
\textcolor{mygreen}{
Given a data reservoir with a fixed memory size \(M\) initially containing the base data block \(D\), the arrival of a new data block \(D'\) of size \(M'\) triggers the sampling of a subset \(\hat{D}\) from the reservoir, sized \(M - M'\). This subset is then combined with \(D'\) to update the reservoir as \(D = \hat{D} \cup D'\). This approach ensures the model is trained on a balanced mix of recent and historical data, effectively mitigating catastrophic forgetting. The sampling process for \(\hat{D}\) can be further optimized using techniques such as random selection or prioritization based on data importance or contribution to the model.
}

\subsection{Loss Change}

\textcolor{mygreen}{
Quantifying the variation in the optimized model's loss upon the exclusion of a single sample from the training set at the onset of training is a pivotal concept, referred to as Loss Change (as defined in Definition~\ref{def:LossChange}). As a metric that quantify a specific training sample on model training \cite{influence_function}, it is potential to provide an avenue in denoising \cite{data_cleansing} and redundant subset removal\cite{dataset_pruning}. In this study, the term Loss Change specifically denotes the differential loss measured on the training dataset
}

\begin{definition}[Loss Change] 
\label{def:LossChange}
Consider a model $f(x,\theta)$ with parameters $\theta\in \mathbf{\Theta}$ that maps input $x\in \mathcal{X}$ to output $y\in \mathcal{Y}$. Given a training dataset $D_{train}=\{z_i=(x_i,y_i)\in \mathcal{X} \times \mathcal{Y}\}$. Let $L$ be the loss function that quantifies the discrepancy between the prediction of $f(x,\theta)$ and the actual label $y$. The prediction loss for $z_i$ is $L(z_i,\theta)=L(f(x_i,\theta),y_i)$, and the overall loss on $D_{train}$ is $L(D_{train},\theta)=\frac{1}{n}\sum^n_{i=1}L(z_i,\theta)$. The deep learning objective is to obtain an optimal model with $\hat{\theta}=\operatorname{argmin_{\theta\in \mathbf{\Theta}}} L(D_{train},\theta)$ that reaches the minimum train loss.
Suppose that a single sample $z_k=(x_k,y_k)$ has been removed from $D_{train}$ and gets a modified dataset, denoted as $D_{k}=D_{train}\setminus z_k$. Train model $f$ on this modified dataset until convergence, and denote the optimal parameters of the resulting model by $\hat{\theta_{k}}=\operatorname{argmin_{\theta\in \mathbf{\Theta}}} L(D_{train}\setminus z_k,\theta)$. The Loss Change for $z_k$ is $\Delta L_{k}=L(D_{train},\hat{\theta_{k}})-L(D_{train},\hat{\theta})$.
\end{definition}



\textcolor{mygreen}{
Investigating Loss Change $\Delta L_{k}$ through retraining presents a formidable undertaking. The SGD-Influence Estimator \cite{data_cleansing} offers an approach to address this issue. By assuming that a deep learning model's optimization algorithm follows stochastic gradient descent (SGD), this method accurately quantifies loss change caused by the exclusion of a specific sample throughout the entire SGD optimization process. The primary limitation of this approach lies in its reliance on storing parameters updated during each iteration, which imposes a significant storage overhead. Furthermore, the method demands considerable computational resources, rendering it inefficient and infeasible for real-world recommender systems that typically involve large-scale datasets and extensive model parameters. These constraints highlight the pressing need for a more streamlined and computationally efficient influence estimation method, one that can better align with the scalability and performance requirements of practical recommender systems. A previous study \cite{interable_debugging} proposed an approach that simplifies the computation of influence estimation by assuming that updated model parameters are derived from a single gradient descent step applied to the original model. This assumption significantly reduces computational complexity while maintaining reasonable accuracy. Building upon this insight, we introduce the concept of \textit{One Step Loss Change}, which quantifies the change in loss occurring during a single optimization step after the removal of a sample. This metric serves as a practical and efficient compromise for estimating loss change, striking a balance between computational efficiency and real-world applicability.
}


\begin{figure}[t]
  \centering
    \includegraphics[width=3.3in]{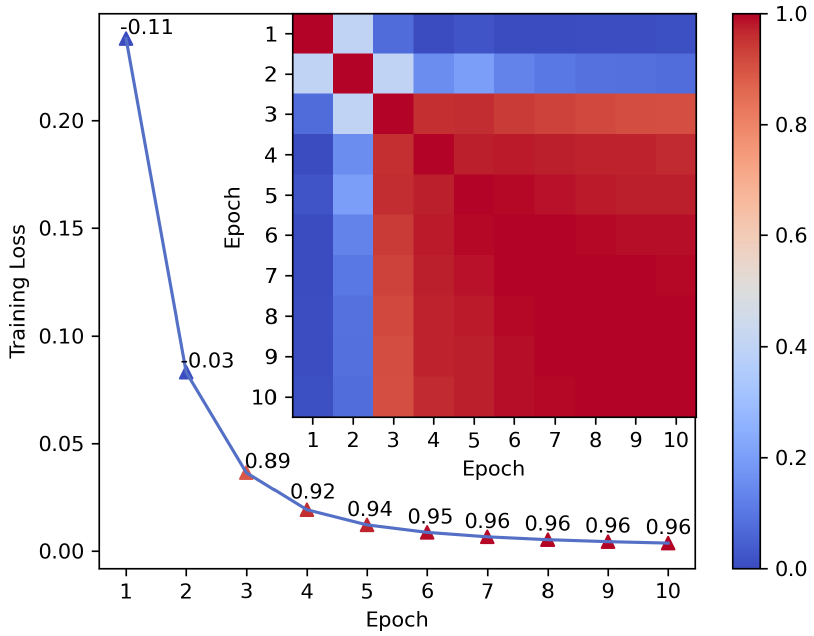}
    \vspace{-2mm}
   \caption{Pearson Correlation Analysis: The line graph depicts the training loss throughout the training process. The annotation showcases the Pearson Correlation between the One Step Loss Change and the Loss Change. Additionally, the heatmap visualizes the Pearson Correlation between each pair of One Step Loss Changes.}
    \label{fig:TrainLossChange_CORR}
\end{figure}


\begin{definition}[One Step Loss Change]
Given an optimization step $i$ where the model parameter is $\theta^i$ and a sample $z_{k}\in D_{train}$. $\theta^{i+1}$ refers to the model parameters obtained by updating $\theta^i$ with one optimization step on $D_{train}$, and $\theta_{k}^{i+1}$ refers to the model parameters obtained by updating $\theta^i$ with identical one optimization step except for the dataset is on $D_{k}$. The One Step Loss Change at step $i$ of a sample $z_{k}\in D_{train}$ is defined as: 

\begin{equation} \label{eq:One Step Loss Change}
\Delta L_{k}^{i}=L(D_{train},\theta_{k}^{i+1})-L(D_{train},\theta^{i+1})
\end{equation}
\end{definition}

To further validate the effectiveness of One Step Loss Change as a practical vanilla Loss Change, we conducted a simple yet insightful empirical study on the MNIST dataset \cite{MNIST}. The study focused on two classes (1 and 7) and utilized a randomly sampled subset of 200 instances. Following prior research \cite{data_cleansing}, we employed a straightforward two-layer neural network trained over 10 epochs. For each training sample, we leveraged the SGD-Influence Estimator to compute 10 One Step Loss Change values at the end of each epoch, alongside the Loss Change across all 10 epochs. To analyze the consistency and reliability of our proposed metric, we calculated the pairwise Pearson correlation coefficients among the 10 sets of One Step Loss Change values. Additionally, we examined the Pearson correlation between each set of One Step Loss Change values and the cumulative Loss Change. The results, visualized in Fig.~\ref{fig:TrainLossChange_CORR}, reveal a striking pattern: as the model approached convergence—most notably from the third epoch onward—the One Step Loss Change demonstrated a strong and increasingly positive correlation with the Loss Change. 
These findings provide compelling empirical support for the utility of One Step Loss Change as an efficient and robust surrogate for the Loss Change, reinforcing the theoretical insights posited in our earlier discussions.

\section{Methodology}
In this section, we start with estimating One Step Loss Change in Section~\ref{sec:theo}, and find that the One Step Loss Change is measured by the dot product between gradients. To further simplify the estimation, in Section~\ref{sec:GGscore}, we introduce a criterion called GGscore to establish a proportional relationship with One Step Loss Change. Subsequently, in Section~\ref{sec:parameter_selection}, we introduced the Model Parameter Selection method to enhance the computational efficiency of GGscore. Lastly in Section~\ref{sec:replay_MEGG}, we proposed our incremental learning method: replay samples with Maximally Extreme GGscore (MEGG). 

\subsection{Estimation of One Step Loss Change}
\label{sec:theo}
We set the optimization method as Mini-Batch Gradient Descent (MBGD), the initial model parameters are denoted as $\theta^0$, and the 0-th batch data is denoted as $\mathbf{Z}^0$. The constant learning rate is symbolized by $\alpha$, and the batch size is $B$. Let $g(D,\theta)=\nabla_\theta L(D,\theta)$ denote the gradient of the loss function with respect to the model parameter $\theta$ on dataset $D$. Hence, the 0-th MBGD optimization step is as follows:

\begin{equation} \label{eq:0-th MBGD}
\theta^{1}=\theta^{0}-\frac{\alpha}{B}\sum _{z\in \mathbf{Z}^{0}}g(z,\theta^0)
\end{equation}

And for the (i-1)-th MBGD optimization step:

\begin{equation} \label{eq:i-th MBGD}
\theta^{i}=\theta^{i-1}-\frac{\alpha}{B}\sum _{z\in {Z}^{i-1}}g(z,\theta^{i-1})
\end{equation}

 Consider the scenario where a single example $z_k$ is removed from the batch ${Z}^{i-1}$, denoted as ${Z}^{i-1}_{k}={Z}^{i-1}\setminus z_{k}$. In this case, the MBGD optimization step will update $\theta^{i-1}$ to $\theta^{i}_{k}$:

\begin{equation} \label{eq:i-th CounterMBGD}
\theta^{i}_{k}=\theta^{i-1}-\frac{\alpha}{B-1}\sum_{z\in {Z}^{i-1}_{k} }g(z,\theta^{i-1})
\end{equation}


\textcolor{mygreen}{%
Let \eqref{eq:i-th CounterMBGD} - \eqref{eq:i-th MBGD}:
}
\begin{equation}\label{formula: model change}
\begin{aligned}
&\quad \theta^{i}_{k}-\theta^{i} \\
&= \frac{\alpha}{B}\sum_{z\in Z^{i-1}} g(z,\theta^{i-1})
   - \frac{\alpha}{B-1}\sum_{z\in Z^{i-1}_{k}} g(z,\theta^{i-1}) \\
&= \frac{\alpha}{B}\sum_{z\in Z^{i-1}} g(z,\theta^{i-1}) \\
&\quad - \frac{\alpha}{B-1}
   \Big( \sum_{z\in Z^{i-1}} g(z,\theta^{i-1}) - g(z_k,\theta^{i-1}) \Big) \\
&= \frac{\alpha}{B}\sum_{z\in Z^{i-1}} g(z,\theta^{i-1}) \\
&\quad - \frac{\alpha}{B-1}\sum_{z\in Z^{i-1}} g(z,\theta^{i-1})
   + \frac{\alpha}{B-1} g(z_k,\theta^{i-1}) \\
&= \frac{\alpha}{B-1} g(z_k,\theta^{i-1}) \\
&\quad + \left(\frac{\alpha}{B}-\frac{\alpha}{B-1}\right)
   \sum_{z\in Z^{i-1}} g(z,\theta^{i-1}) \\
&= \frac{\alpha}{B-1} g(z_k,\theta^{i-1}) - \frac{\alpha}{B-1}\cdot \frac{1}{B}
   \sum_{z\in Z^{i-1}} g(z,\theta^{i-1}) \\
&= \frac{\alpha}{B-1}\big( g(z_k,\theta^{i-1})
     - g(Z^{i-1},\theta^{i-1}) \big)
\end{aligned}
\end{equation}


where $g({Z}^{i-1}, \theta^{i-1}) = \frac{1}{B} \sum_{z \in {Z}^{i-1}} g(z, \theta^{i-1})$, for $L(D_{train},\theta^{i+1}_{k})$, its first-order Taylor expansion at $\theta^{i+1}$ is as follows:
\begin{equation} 
\begin{aligned}
L(D_{train}, &\theta^{i+1}_{k}) = L(D_{train}, \theta^{i+1})  \\ & + \nabla_{\theta^{i+1}}L(D_{train},\theta^{i+1}) \cdot (\theta^{i+1}_{k} - \theta^{i+1})  \\ & + o(\| \theta^{i+1}_{k} - \theta^{i+1} \|)
\end{aligned}
\end{equation}

Under the assumption that the parameter change $\theta^{i+1}_{k} - \theta^{i+1}$ is minimal, the infinitesimal term $o$ can be neglected. Then we can quantify the One Step Loss Change in the following manner:

\begin{equation} \label{eq: loss change}
\begin{aligned}
 \Delta L_k^{i} 
&=L(D_{train}, \theta^{i+1}_{k})-L(D_{train}, \theta^{i+1}) \\
&\approx \nabla_{\theta^{i+1}}L(D_{train},\theta^{i+1})\cdot(\theta^{i+1}_{k}-\theta^{i+1}) \\
&\approx \frac{\alpha}{B-1}  g(D_{train},\theta^{i+1})\cdot(g(z_k,\theta^{i}) - g({Z}^{i},\theta^{i}) ) \\
\end{aligned}
\end{equation}

\subsection{GGscore}
\label{sec:GGscore}
\eqref{eq: loss change} reveals that the One Step Loss Change in MBGD training can be measured by the dot product between gradients. Inspired by this, we define a metric called GGscore as in Definition~(\ref{def:GGscore}). 

\begin{definition}[GGscore]
\label{def:GGscore}
Given a sample $z$ and model parameter $\theta$, and a reference vector $\mathbf{V}\subseteq \mathbb{R}^{|\theta|}$. Then the GGscore of $z$ is defined as: 
\begin{equation} \label{eq:GG Score}
GG(z,\theta \mid \mathbf{V})=\mathbf{V} \cdot \nabla_{\theta} L(z,\theta)
\end{equation}
\end{definition}
The GGscore is applicable to wide deep-learning settings. Literally, the GGscore quantifies the alignment between the gradients of a given sample $z$ with respect to the model parameters $\theta$ and a designated reference vector $\mathbf{V}$ of the same dimensionality. In a physical sense, the GGscore can be interpreted as a measure of the influence of individual samples on model training: The reference vector $\mathbf{V}$ indicates the subsequent direction of model convergence, and the parameter $\theta$ signifies the current influence of sample $z$ on model training when the parameters are $\theta$.

Our investigation has revealed that under certain specific conditions,  One Step Loss Change presents a proportional relation to GGscore in MBGD. Let $\theta^{i}$ denote the model parameters, and $\theta^{i+1}$ represent the updated parameters after performing a MBGD optimization step. Let $\mathbf{V}=\nabla_{\theta^{i+1}}L(D_{train},\theta^{i+1})$, then $\forall z_{k}\in D_{train}$, the One Step Loss Change $\Delta L_k^i$ and GGscore $GG(z_{k},\theta^i | \nabla_{\theta^{i+1}}L(D_{train},\theta^{i+1}))$ satisfy Equation~\eqref{eq:proportional}:
\begin{equation} \label{eq:proportional}
 \Delta L_k^i \propto GG(z_{k},\theta^i | \nabla_{\theta^{i+1}}L(D_{train},\theta^{i+1}))
\end{equation}


The assertion that the One-Step Loss Change is proportional to the GGScore can be derived with relative simplicity. This conclusion is reached by omitting the constant term \(\frac{\alpha}{B-1}\) in Equation \(\eqref{eq: loss change}\) and disregarding the interaction term \(g(D_{train}, \theta^{i+1}) \cdot g({Z}^{i}, \theta^{i})\). A critical observation here is that the per-sample loss change, \(\Delta L_k^{i}\), for a given sample \(z_k\), is inherently influenced by the composition of its corresponding mini-batch, \({Z}^{i}\), during the Mini-Batch Gradient Descent (MBGD) process. This stochasticity, intrinsic to deep learning, introduces noise that can obscure the investigation into the fundamental influence properties of individual samples. To mitigate this, we propose a modification to Equation \(\eqref{eq:i-th CounterMBGD}\), replacing \(B-1\) with \(B\). This adjustment effectively eliminates the term \(g({Z}^{i}, \theta^{i})\) from \(\Delta L_k^{i}\) in Equation \(\eqref{eq: loss change}\). While this alteration introduces a minor departure from the conventional parameter update mechanism in MBGD, it significantly enhances the clarity and tractability of our analysis of influence.

\subsection{Model Parameter Selection}
\label{sec:parameter_selection}
In this section, we present parameter selection techniques aimed at improving the efficiency of computing GGscore. As delineated in Equation~\eqref{eq:GG Score}, per-sample gradients need to be computed, while this task is nearly unattainable for large-scale recommendation models. Prior research has shed light on the fact that it is not imperative to utilize all model parameters to measure the influence. For example, empirical studies in \cite{selective_rec_unlearning} demonstrate that removing one user's interaction records primarily affects the embedding of the target users, while having minimal impact on the embeddings of other users and remaining parameters. Moreover, in \cite{dataset_pruning}, the Influence Function is computed by considering only the parameters of the final fully connected layer of a model. Building on these foundational insights, we adopt a similar approach. Specifically, for computing the GGscore of a sample, we only calculate the gradients with respect to the embedding parameters associated with the user and the item in an interaction record, as well as the parameters of the final fully connected layer. The gradients for all other parameters are set to zero. 

\subsection{Replay via Maximally Extreme GGscore}
\label{sec:replay_MEGG}

In this subsection, we present our proposed incremental learning method: replaying samples with Maximally Extreme GGscore (MEGG). 
The process involves the following steps, and we repeat the step 1-4 periodically in the online dynamic recommender systems. A detailed description of our proposed method can be found in Algorithm~\ref{alg:algorithm}.

Step 1: We start with a reservoir that stores historical data $D$ of size $M$. A deep learning model $f$ is trained from scratch on $D$ until convergence is achieved. During the training process, we maintain records of two sets of model parameters. One is the parameters $\hat{\theta}$ at the point of convergence, for instance, the parameters after the final epoch. The other is the parameters $\hat{\theta}^{'}$ before convergence, such as the parameters after the penultimate epoch. Further robustness experiments regarding the selection of $\hat{\theta}$ and $\hat{\theta}^{'}$ can be found in Section~\ref{sec: hyperparameter_analysis}.

Step 2: According to Equation~\eqref{eq:GG Score}, let the reference vector $\mathbf{V}=\nabla_{\hat{\theta}}L(D,\hat{\theta})$ and $\theta=\hat{\theta}^{'}$, then compute the GGscore $GG(z_k,\hat{\theta}^{'} \mid \nabla_{\hat{\theta}}L(D,\hat{\theta})$ of each sample $z_{k} \in D$.



Step 3: Let
$
s_k \triangleq GG\!\left(z_k,\hat{\theta}^{'} \mid g(D,\hat{\theta})\right).
$
Sort samples in non-decreasing order of scores and denote the ordered pairs by
$\bigl(z_{(1)},s_{(1)}\bigr),\ \bigl(z_{(2)},s_{(2)}\bigr),\ \ldots,\ \bigl(z_{(|D|)},s_{(|D|)}\bigr)$, with $s_{(1)}\le \cdots \le s_{(|D|)}$. Let the required memory size be \(K \triangleq M-M'\) and define
$
k_\ell \triangleq \left\lfloor \frac{K}{2}\right\rfloor,
k_r \triangleq \left\lceil \frac{K}{2}\right\rceil .
$
The “greedy keep-both-ends” subset is then
\begin{equation}\label{eq:hatD_sort}
\begin{aligned}
\hat{D}_{\mathrm{left}}  &= \{\, z_{(i)} \;:\; 1 \le i \le k_\ell \,\},\\
\hat{D}_{\mathrm{right}} &= \{\, z_{(i)} \;:\; |D|-k_r+1 \le i \le |D| \,\},\\[2pt]
\hat{D} &= \hat{D}_{\mathrm{left}} \cup \hat{D}_{\mathrm{right}}.
\end{aligned}
\end{equation}

Step 4: We merge the retained data $\hat{D}$ with the incremental data $D'$ to update the reservoir $D = \hat{D} \cup D'$.

\begin{algorithm}
 \caption{MEGG}
 \label{alg:algorithm}
 
 \KwIn{A reservoir stores historical data $D$ of size $M$. A randomly initialized deep learning model $f$.}
 \textbf{Process}:\;
\Repeat
        {\text{service end}}
{        
Train deep learning model $f$ from scratch on $D$ until convergence is achieved. The model parameter at the point of convergence is denoted as $\hat{\theta}$, and the model parameter trained after the penultimate epoch is denoted as $\hat{\theta}^{'}$\;

Compute the GGscore of each sample $z_{k} \in D$ with the reference vector $\mathbf{V}=\nabla_{\hat{\theta}}L(D,\hat{\theta})$ and $\theta=\hat{\theta}^{'}$ with \eqref{eq:GG Score};

Select samples with maximally extreme GGscore to construct the subset of data $\hat{D}\subseteq D$ with \eqref{eq: hat_D};

Update the reservoir $D$ by merging $\hat{D}$ with the incremental data $D'$: $D=\hat{D}\cup D^{'}$\;
}

\end{algorithm}

\section{Experiments}
In this section, we take both the rating prediction and classification prediction as recommendation tasks. We present extensive experiments to evaluate our proposed method MEGG and make comparisons with other state-of-the-art baseline methods. Our primary focus is on addressing the following research questions:

• \textbf{RQ1}: What is the performance of MEGG compared with existing experience replay-based incremental learning methods, and compared with other category incremental recommendation methods?

• \textbf{RQ2}: Compared to not combining, what are the performance enhancements of incremental recommendation methods when combined with MEGG? Additionally, how does this improvement compare to the enhancement obtained through the combination with other replay methods?

• \textbf{RQ3}: How does the sampling efficiency of the MEGG method vary across different embedding sizes, and how does it stand in comparison to other sampling techniques?

• \textbf{RQ4}: How do the hyper-parameters $\hat{\theta}^{'}$, $\hat{\theta}$, and $\mathbf{V}$ in Algorithm~\ref{alg:algorithm} affect the effectiveness of MEGG?  

• \textbf{RQ5}: How does the replay ratio affect the effectiveness of MEGG?

\subsection{Dataset}
\textcolor{mygreen}{
We conducted extensive experiments on four benchmark datasets: MovieLens 1M, Douban Movie \cite{FIRE}, Taobao2014, and LastFM-1k \cite{lastfm-1k}, all of which have been widely adopted in prior research \cite{graphsail, FIRE, graph_structure_aware_replay, fast_inc_FM}. These datasets exhibit significant diversity in terms of size, sparsity, and temporal dynamics, making them highly suitable for modeling and evaluating online dynamic recommender systems across varied domains and temporal scales. Notably, the first three datasets provide explicit user ratings, while the latter focuses on implicit user interactions—a common characteristic of real-world recommendation scenarios. By incorporating both explicit and implicit feedback, our experimental setup ensures a comprehensive evaluation of the proposed methods.
}

\textbf{MovieLens 1M}\footnote{https://grouplens.org/datasets/movielens/1m/}: MovieLens 1M dataset is a widely used benchmark dataset in the field of recommender systems. It consists of 1 million movie ratings provided by 6,000 users on 4,000 movies. Each rating ranges from 1 to 5 for rating prediction. For classification, the ratings of 1 and 2 are considered as negative samples, and ratings of 4 and 5 are considered as positive samples.

\textbf{Douban Movie}\footnote{https://github.com/Yaveng/FIRE/tree/main/dataset}: The Douban Movie dataset consists of 1 million movie ratings, which were collected from the Douban website and ranged between 1 to 5 for rating prediction. The dataset includes data from 10,000 users and 10,000 movies spanning the years 2008 to 2019. The settings for classification is the same as in MovieLens-1M. 

\textbf{LastFM-1k}\footnote{http://ocelma.net/MusicRecommendationDataset/lastfm-1K.html}:
The LastFM-1k dataset is a music listening history dataset consisting of data from nearly 992 users. The dataset includes 17 million records, each containing information on the user, timestamp, artist, and song. We took the earliest year's (2005) worth of data from whole dataset for the experiment. To apply this dataset for rating prediction, we label the user's rating toward a song via record frequency as follows: 5 for listening to a song 30 or more times, 4 for 15 to 29 times, 3 for 5 to 14 times, 2 for 1 to 4 times, and 1 for 0 times. The settings for classification is the same as in MovieLens-1M.

\textbf{Taobao2014}\footnote{https://tianchi.aliyun.com/dataset/46}:
The Taobao2014 dataset contains mobile behavior data from 10k users on nearly 3 million items, collected from November 18th to December 19th, 2014. The records include information on user, item, category, time, and behavior type (Including click, collect, add-to-cart, and payment, corresponding to the rating value of 1, 2, 3 and 4, respectively.). We took all the interaction records of 1,000 random users from the dataset for our experiments. For classification, the ratings of 1 and 2 are considered as negative samples, and ratings of 3 and 4 are considered as positive samples.


\subsection{Experiment setting}
\textcolor{mygreen}{
To align with previous studies \cite{FIRE,huawei_IRS,ADER,graph_structure_aware_replay} and simulate real-world scenarios, we partition the dataset into 15 equal-sized blocks in chronological order. The first 10 blocks are used to form the initial reservoir \(D\), representing the system's training data after a period of operation. The size of the reservoir is set to \(M = \frac{10}{15}\) of the total dataset, which corresponds to the system's maximum capacity for storing historical data. The remaining 5 blocks, each constituting \(M' = \frac{1}{15}\) of the total dataset, serve as incremental data blocks \(D'\), representing newly generated user-item interactions over successive time periods. The online servicing process is organized into 5 continuous stages, with each stage corresponding to one incremental data block \(D'\). After each stage, the model is updated using both the new data and the historical data stored in the reservoir. This process simulates the continuous learning scenario in which the system adapts to new user interactions while retaining past knowledge to avoid catastrophic forgetting. Specifically, during each stage, the Experience Replay procedure is employed to sample a subset \(\hat{D}\) from the reservoir and combine it with the incoming data block \(D'\).
}

\textbf{Evaluation Metrics}:
After a stage, the model is updated and the performance is evaluated on the remaining incremental data blocks. We use metric RMSE to evaluate the rating prediction performance and use metric AUC to evaluate the classification performance. We calculate the average RMSE, and average AUC across stages 1 to 4 as the metric for evaluation. It should be noticed that, an improvement of RMSE and AUC at 0.1\% can be considered as significant for a recommendation model \cite{huawei_IRS}.

 
\textbf{Backbone Models}:
Our MEGG approach is versatile and can be applied to a broad range of neural network-based collaborative filtering (NCF) \cite{NCF} models. In this study, we utilize three popular models as the backbone methods.

1.Wide \& Deep \cite{wide_deep} combines the strengths of deep neural networks and traditional linear models to provide high accuracy and generalization in large-scale recommender systems. 

2.DCN \cite{DCN} combines the advantages of deep neural networks with a novel cross-network, enabling efficient learning of feature interactions.

3.NFM \cite{neural_FM} combines
the linearity of factorization machines and non-linearity of the neural network to achieve
recommendation.

\textbf{Baseline Algorithms}: 
Basically, we regard the Full-Batch and Fine-Tune methods as the upper and lower bounds, respectively. To frame our comparison within the context of experience replay-based methods, we select three representative sampling strategies exist in iCaRL, MIR, and GDumb, respectively. Since these methods were originally designed for class incremental learning task, we have omitted the class balance part to adapt them to the recommendation settings. Furthermore, we also compare with other category incremental learning methods for neural recommendation models, including IncCTR and SML, to extend our comparison to a broader scope.

1.Full-Batch: This method involves training the model from scratch at each incremental stage using all the historical data.

2.Fine-Tune: This naive baseline involves fine-tuning the previously trained model on the incremental block.

3.iCaRL\cite{iCaRL}: This method advocates for preserving samples that are closest to the feature center. We regard the input of the last fully connected layer as the feature representation.

4.MIR\cite{MIR}: This method advocates for preserving samples near the decision boundary, it retrieves the samples whose prediction will be most negatively impacted by the foreseen parameters update.

5.GDumb\cite{Gdumb}: This method randomly samples historical records, it serves as a strong and robust baseline surpassing many existing repay methods. 

6.IncCTR\cite{huawei_IRS}: This framework enables the incremental deployment of Deep CTR models in practical recommendation systems. The method primarily utilizes the knowledge distillation technique to address the issue of forgetting. Although it also proposes a data module for replaying historically valuable samples, this aspect of the work remains conceptual.

7.SML\cite{SML}: This is a state-of-the-art incremental recommendation method that employs a neural network-based meta-learning component to acquire a model through learning from an old model and a new model trained on the incremental data block.

\textbf{Implementation Details}: Our experiments utilized Python 3.8, PyTorch 1.13, CUDA 11.6, and a single Quadro RTX 6000 GPU. We set the embedding dimensions as 64 for all ID-type data across four datasets. During the training phase, the model parameters are optimized by the Adam optimizer with a fixed learning rate of 0.001 for all experiments. In each stage, we train models for a total of 5 epochs with a batch size of 1024. To efficiently calculate the GGscore for all samples, we adopted the VMAP technique~ \footnote{https://pytorch.org/tutorials/intermediate/per\_sample\_grads.html} to vectorize the calculation of GGscore over multiple independent tensors. All experiments are conducted with 5 different random seeds and we report the mean value and standard deviation. Additionally, the implementation of SML follows \footnote{https://github.com/zyang1580/SML}, and the hyperparameter for the maximum number of iterations is tuned within the range $\{5, 6, \cdots, 10\}$, in alignment with previous work \cite{SML}.

{
\setlength{\tabcolsep}{1pt}
\begin{table*}[t] 
\fontsize{7pt}{8.6pt}\selectfont 

\caption{Recommendation performance comparison between MEGG and other methods in four datasets using three models.}
\label{tab:result1}
\centering
\vspace{-0.2cm}
\begin{tabular}{cc|cccccccccccc}
\hline
\multirow{2}{*}{Metric} & \multirow{2}{*}{Method} & \multicolumn{3}{c|}{MovieLens-1M} & \multicolumn{3}{c|}{Douban} & \multicolumn{3}{c|}{LastFM-1K} & \multicolumn{3}{c}{Taobao204} \\ \cline{3-14} 
 &  & WDL & DCN & \multicolumn{1}{c|}{NFM} & WDL & DCN & \multicolumn{1}{c|}{NFM} & WDL & DCN & \multicolumn{1}{c|}{NFM} & WDL & DCN & NFM \\ \hline
\multirow{14}{*}{\begin{tabular}[c]{@{}c@{}}AVG \\ RMSE\end{tabular}} & FT & \begin{tabular}[c]{@{}c@{}}0.9657\\ ±0.0010\end{tabular} & \begin{tabular}[c]{@{}c@{}}0.9506\\ ±0.0005\end{tabular} & \begin{tabular}[c]{@{}c@{}}0.9850\\ ±0.0023\end{tabular} & \begin{tabular}[c]{@{}c@{}}0.8530\\ ±0.0012\end{tabular} & \begin{tabular}[c]{@{}c@{}}0.8690\\ ±0.0011\end{tabular} & \begin{tabular}[c]{@{}c@{}}0.8907\\ ±0.0019\end{tabular} & \begin{tabular}[c]{@{}c@{}}0.8862\\ ±0.0008\end{tabular} & \begin{tabular}[c]{@{}c@{}}0.8291\\ ±0.0007\end{tabular} & \begin{tabular}[c]{@{}c@{}}0.7618\\ ±0.0016\end{tabular} & \begin{tabular}[c]{@{}c@{}}0.4972\\ ±0.0003\end{tabular} & \begin{tabular}[c]{@{}c@{}}0.4952\\ ±0.0004\end{tabular} & \begin{tabular}[c]{@{}c@{}}0.4905\\ ±0.0003\end{tabular} \\
 & iCaRL & \begin{tabular}[c]{@{}c@{}}0.9582\\ ±0.0013\end{tabular} & \begin{tabular}[c]{@{}c@{}}0.9427\\ ±0.0002\end{tabular} & \begin{tabular}[c]{@{}c@{}}0.9395\\ ±0.0012\end{tabular} & \begin{tabular}[c]{@{}c@{}}0.8379\\ ±0.0005\end{tabular} & \begin{tabular}[c]{@{}c@{}}0.8391\\ ±0.0010\end{tabular} & \begin{tabular}[c]{@{}c@{}}0.8490\\ ±0.0008\end{tabular} & \begin{tabular}[c]{@{}c@{}}0.8757\\ ±0.0008\end{tabular} & \begin{tabular}[c]{@{}c@{}}0.8184\\ ±0.0007\end{tabular} & \begin{tabular}[c]{@{}c@{}}0.7694\\ ±0.0003\end{tabular} & \begin{tabular}[c]{@{}c@{}}0.4886\\ ±0.0003\end{tabular} & \begin{tabular}[c]{@{}c@{}}0.4905\\ ±0.0005\end{tabular} & \begin{tabular}[c]{@{}c@{}}0.4880\\ ±0.0001\end{tabular} \\
 & MIR & \begin{tabular}[c]{@{}c@{}}0.9668\\ ±0.0024\end{tabular} & \begin{tabular}[c]{@{}c@{}}0.9719\\ ±0.0045\end{tabular} & \begin{tabular}[c]{@{}c@{}}0.9480\\ ±0.0023\end{tabular} & \begin{tabular}[c]{@{}c@{}}0.8415\\ ±0.0006\end{tabular} & \begin{tabular}[c]{@{}c@{}}0.8587\\ ±0.0029\end{tabular} & \begin{tabular}[c]{@{}c@{}}0.8550\\ ±0.0015\end{tabular} & \begin{tabular}[c]{@{}c@{}}0.868\\ ±0.0013\end{tabular} & \begin{tabular}[c]{@{}c@{}}0.841\\ ±0.0021\end{tabular} & \begin{tabular}[c]{@{}c@{}}0.7516\\ ±0.0016\end{tabular} & \begin{tabular}[c]{@{}c@{}}0.5083\\ ±0.0046\end{tabular} & \begin{tabular}[c]{@{}c@{}}0.5452\\ ±0.0082\end{tabular} & \begin{tabular}[c]{@{}c@{}}0.4897\\ ±0.0005\end{tabular} \\
 & GDumb & \begin{tabular}[c]{@{}c@{}}0.9554\\ ±0.0012\end{tabular} & \begin{tabular}[c]{@{}c@{}}0.9413\\ ±0.0011\end{tabular} & \begin{tabular}[c]{@{}c@{}}0.9363\\ ±0.0012\end{tabular} & \begin{tabular}[c]{@{}c@{}}0.8380\\ ±0.0002\end{tabular} & \begin{tabular}[c]{@{}c@{}}0.8389\\ ±0.0009\end{tabular} & \begin{tabular}[c]{@{}c@{}}0.8488\\ ±0.0017\end{tabular} & \begin{tabular}[c]{@{}c@{}}0.8642\\ ±0.0003\end{tabular} & \textbf{\begin{tabular}[c]{@{}c@{}}0.8018\\ ±0.0011\end{tabular}} & \begin{tabular}[c]{@{}c@{}}0.7566\\ ±0.0021\end{tabular} & \begin{tabular}[c]{@{}c@{}}0.4888\\ ±0.0004\end{tabular} & \textbf{\begin{tabular}[c]{@{}c@{}}0.4895\\ ±0.0003\end{tabular}} & \begin{tabular}[c]{@{}c@{}}0.4878\\ ±0.0001\end{tabular} \\
 & IncCTR & \begin{tabular}[c]{@{}c@{}}0.9568\\ ±0.0005\end{tabular} & \begin{tabular}[c]{@{}c@{}}0.9391\\ ±0.0011\end{tabular} & \begin{tabular}[c]{@{}c@{}}0.9388\\ ±0.0017\end{tabular} & \begin{tabular}[c]{@{}c@{}}0.8413\\ ±0.0005\end{tabular} & \begin{tabular}[c]{@{}c@{}}0.8465\\ ±0.0005\end{tabular} & \begin{tabular}[c]{@{}c@{}}0.8531\\ ±0.0005\end{tabular} & \begin{tabular}[c]{@{}c@{}}0.8791\\ ±0.0006\end{tabular} & \begin{tabular}[c]{@{}c@{}}0.8144\\ ±0.0008\end{tabular} & \begin{tabular}[c]{@{}c@{}}0.7578\\ ±0.0010\end{tabular} & \begin{tabular}[c]{@{}c@{}}0.4893\\ ±0.0003\end{tabular} & \begin{tabular}[c]{@{}c@{}}0.4912\\ ±0.0010\end{tabular} & \begin{tabular}[c]{@{}c@{}}0.4884\\ ±0.0003\end{tabular} \\
 & SML & \begin{tabular}[c]{@{}c@{}}0.9548\\ ±0.0002\end{tabular} & \begin{tabular}[c]{@{}c@{}}0.9401\\ ±0.0008\end{tabular} & \begin{tabular}[c]{@{}c@{}}0.9354\\ ±0.0023\end{tabular} & \begin{tabular}[c]{@{}c@{}}0.8453\\ ±0.0008\end{tabular} & \begin{tabular}[c]{@{}c@{}}0.8434\\ ±0.0011\end{tabular} & \begin{tabular}[c]{@{}c@{}}0.8512\\ ±0.0008\end{tabular} & \begin{tabular}[c]{@{}c@{}}0.8638\\ ±0.0004\end{tabular} & \begin{tabular}[c]{@{}c@{}}0.8077\\ ±0.0011\end{tabular} & \begin{tabular}[c]{@{}c@{}}0.7526\\ ±0.0008\end{tabular} & \textbf{\begin{tabular}[c]{@{}c@{}}0.4878\\ ±0.0003\end{tabular}} & \begin{tabular}[c]{@{}c@{}}0.4903\\ ±0.0010\end{tabular} & \begin{tabular}[c]{@{}c@{}}0.4872\\ ±0.0001\end{tabular} \\
 & MEGG & \textbf{\begin{tabular}[c]{@{}c@{}}0.9530\\ ±0.0013\end{tabular}} & \textbf{\begin{tabular}[c]{@{}c@{}}0.9378\\ ±0.0007\end{tabular}} & \textbf{\begin{tabular}[c]{@{}c@{}}0.9326\\ ±0.0026\end{tabular}} & \textbf{\begin{tabular}[c]{@{}c@{}}0.8334\\ ±0.0003\end{tabular}} & \textbf{\begin{tabular}[c]{@{}c@{}}0.8351\\ ±0.0015\end{tabular}} & \textbf{\begin{tabular}[c]{@{}c@{}}0.8456\\ ±0.0012\end{tabular}} & \textbf{\begin{tabular}[c]{@{}c@{}}0.8626\\ ±0.0003\end{tabular}} & \begin{tabular}[c]{@{}c@{}}0.8026\\ ±0.0004\end{tabular} & \textbf{\begin{tabular}[c]{@{}c@{}}0.7446\\ ±0.0009\end{tabular}} & \begin{tabular}[c]{@{}c@{}}0.4880\\ ±0.0004\end{tabular} & \begin{tabular}[c]{@{}c@{}}0.4905\\ ±0.0008\end{tabular} & \textbf{\begin{tabular}[c]{@{}c@{}}0.4858\\ ±0.0001\end{tabular}} \\ \cline{2-14} 
 & FB & \begin{tabular}[c]{@{}c@{}}0.9512\\ ±0.0008\end{tabular} & \begin{tabular}[c]{@{}c@{}}0.9373\\ ±0.0015\end{tabular} & \begin{tabular}[c]{@{}c@{}}0.9316\\ ±0.0009\end{tabular} & \begin{tabular}[c]{@{}c@{}}0.8368\\ ±0.0003\end{tabular} & \begin{tabular}[c]{@{}c@{}}0.8388\\ ±0.0003\end{tabular} & \begin{tabular}[c]{@{}c@{}}0.8464\\ ±0.0003\end{tabular} & \begin{tabular}[c]{@{}c@{}}0.8641\\ ±0.0006\end{tabular} & \begin{tabular}[c]{@{}c@{}}0.8014\\ ±0.0003\end{tabular} & \begin{tabular}[c]{@{}c@{}}0.7542\\ ±0.0003\end{tabular} & \begin{tabular}[c]{@{}c@{}}0.4874\\ ±0.0002\end{tabular} & \begin{tabular}[c]{@{}c@{}}0.4898\\ ±0.0005\end{tabular} & \begin{tabular}[c]{@{}c@{}}0.4867\\ ±0.0001\end{tabular} \\ \hline
\multirow{14}{*}{\begin{tabular}[c]{@{}c@{}}AVG \\ AUC\end{tabular}} & FT & \begin{tabular}[c]{@{}c@{}}0.8171\\ ±0.0002\end{tabular} & \begin{tabular}[c]{@{}c@{}}0.8278\\ ±0.0003\end{tabular} & \begin{tabular}[c]{@{}c@{}}0.8176\\ ±0.0021\end{tabular} & \begin{tabular}[c]{@{}c@{}}0.8637\\ ±0.0006\end{tabular} & \begin{tabular}[c]{@{}c@{}}0.8570\\ ±0.0001\end{tabular} & \begin{tabular}[c]{@{}c@{}}0.8407\\ ±0.0007\end{tabular} & \begin{tabular}[c]{@{}c@{}}0.8761\\ ±0.0002\end{tabular} & \begin{tabular}[c]{@{}c@{}}0.9136\\ ±0.0001\end{tabular} & \begin{tabular}[c]{@{}c@{}}0.9159\\ ±0.0004\end{tabular} & \begin{tabular}[c]{@{}c@{}}0.6253\\ ±0.0019\end{tabular} & \begin{tabular}[c]{@{}c@{}}0.6295\\ ±0.0027\end{tabular} & \begin{tabular}[c]{@{}c@{}}0.6407\\ ±0.0011\end{tabular} \\
 & iCaRL & \begin{tabular}[c]{@{}c@{}}0.8203\\ ±0.0008\end{tabular} & \begin{tabular}[c]{@{}c@{}}0.8314\\ ±0.0003\end{tabular} & \begin{tabular}[c]{@{}c@{}}0.8362\\ ±0.0008\end{tabular} & \begin{tabular}[c]{@{}c@{}}0.8743\\ ±0.0002\end{tabular} & \begin{tabular}[c]{@{}c@{}}0.8729\\ ±0.0002\end{tabular} & \begin{tabular}[c]{@{}c@{}}0.8641\\ ±0.0009\end{tabular} & \begin{tabular}[c]{@{}c@{}}0.8970\\ ±0.0002\end{tabular} & \begin{tabular}[c]{@{}c@{}}0.9135\\ ±0.0008\end{tabular} & \begin{tabular}[c]{@{}c@{}}0.9254\\ ±0.0002\end{tabular} & \begin{tabular}[c]{@{}c@{}}0.6386\\ ±0.0025\end{tabular} & \begin{tabular}[c]{@{}c@{}}0.6427\\ ±0.0005\end{tabular} & \begin{tabular}[c]{@{}c@{}}0.6459\\ ±0.0016\end{tabular} \\
 & MIR & \begin{tabular}[c]{@{}c@{}}0.8142\\ ±0.0013\end{tabular} & \begin{tabular}[c]{@{}c@{}}0.8188\\ ±0.0022\end{tabular} & \begin{tabular}[c]{@{}c@{}}0.8336\\ ±0.0012\end{tabular} & \begin{tabular}[c]{@{}c@{}}0.8720\\ ±0.0004\end{tabular} & \begin{tabular}[c]{@{}c@{}}0.8607\\ ±0.0007\end{tabular} & \begin{tabular}[c]{@{}c@{}}0.8604\\ ±0.0009\end{tabular} & \begin{tabular}[c]{@{}c@{}}0.8969\\ ±0.0007\end{tabular} & \begin{tabular}[c]{@{}c@{}}0.9077\\ ±0.0015\end{tabular} & \begin{tabular}[c]{@{}c@{}}0.9283\\ ±0.00018\end{tabular} & \begin{tabular}[c]{@{}c@{}}0.5868\\ ±0.0044\end{tabular} & \begin{tabular}[c]{@{}c@{}}0.6024\\ ±0.0074\end{tabular} & \begin{tabular}[c]{@{}c@{}}0.6294\\ ±0.0038\end{tabular} \\
 & GDumb & \begin{tabular}[c]{@{}c@{}}0.8233\\ ±0.0006\end{tabular} & \textbf{\begin{tabular}[c]{@{}c@{}}0.8335\\ ±0.0004\end{tabular}} & \begin{tabular}[c]{@{}c@{}}0.8389\\ ±0.0010\end{tabular} & \begin{tabular}[c]{@{}c@{}}0.8740\\ ±0.0004\end{tabular} & \begin{tabular}[c]{@{}c@{}}0.8722\\ ±0.0004\end{tabular} & \begin{tabular}[c]{@{}c@{}}0.8636\\ ±0.0008\end{tabular} & \begin{tabular}[c]{@{}c@{}}0.8994\\ ±0.0004\end{tabular} & \begin{tabular}[c]{@{}c@{}}0.9155\\ ±0.0002\end{tabular} & \begin{tabular}[c]{@{}c@{}}0.9262\\ ±0.0008\end{tabular} & \begin{tabular}[c]{@{}c@{}}0.6396\\ ±0.0033\end{tabular} & \begin{tabular}[c]{@{}c@{}}0.6339\\ ±0.0015\end{tabular} & \begin{tabular}[c]{@{}c@{}}0.6471\\ ±0.002\end{tabular} \\
 & IncCTR & \begin{tabular}[c]{@{}c@{}}0.8211\\ ±0.0002\end{tabular} & \begin{tabular}[c]{@{}c@{}}0.8326\\ ±0.0005\end{tabular} & \begin{tabular}[c]{@{}c@{}}0.8366\\ ±0.0003\end{tabular} & \begin{tabular}[c]{@{}c@{}}0.8707\\ ±0.0002\end{tabular} & \begin{tabular}[c]{@{}c@{}}0.8680\\ ±0.0001\end{tabular} & \begin{tabular}[c]{@{}c@{}}0.8608\\ ±0.0007\end{tabular} & \begin{tabular}[c]{@{}c@{}}0.9008\\ ±0.0001\end{tabular} & \begin{tabular}[c]{@{}c@{}}0.9132\\ ±0.0001\end{tabular} & \begin{tabular}[c]{@{}c@{}}0.9258\\ ±0.0003\end{tabular} & \begin{tabular}[c]{@{}c@{}}0.624\\ ±0.0018\end{tabular} & \begin{tabular}[c]{@{}c@{}}0.6407\\ ±0.0014\end{tabular} & \begin{tabular}[c]{@{}c@{}}0.6449\\ ±0.0019\end{tabular} \\
 & SML & \begin{tabular}[c]{@{}c@{}}0.8201\\ ±0.0008\end{tabular} & \begin{tabular}[c]{@{}c@{}}0.8329\\ ±0.0008\end{tabular} & \begin{tabular}[c]{@{}c@{}}0.8404\\ ±0.0007\end{tabular} & \begin{tabular}[c]{@{}c@{}}0.8711\\ ±0.0004\end{tabular} & \begin{tabular}[c]{@{}c@{}}0.8713\\ ±0.0002\end{tabular} & \begin{tabular}[c]{@{}c@{}}0.8631\\ ±0.0008\end{tabular} & \textbf{\begin{tabular}[c]{@{}c@{}}0.9014\\ ±0.0003\end{tabular}} & \begin{tabular}[c]{@{}c@{}}0.9125\\ ±0.0004\end{tabular} & \begin{tabular}[c]{@{}c@{}}0.9288\\ ±0.0007\end{tabular} & \textbf{\begin{tabular}[c]{@{}c@{}}0.6431\\ ±0.0018\end{tabular}} & \begin{tabular}[c]{@{}c@{}}0.6402\\ ±0.0015\end{tabular} & \begin{tabular}[c]{@{}c@{}}0.6481\\ ±0.002\end{tabular} \\
 & MEGG & \textbf{\begin{tabular}[c]{@{}c@{}}0.8250\\ ±0.0005\end{tabular}} & \begin{tabular}[c]{@{}c@{}}0.8332\\ ±0.0003\end{tabular} & \textbf{\begin{tabular}[c]{@{}c@{}}0.8457\\ ±0.0006\end{tabular}} & \textbf{\begin{tabular}[c]{@{}c@{}}0.8826\\ ±0.0004\end{tabular}} & \textbf{\begin{tabular}[c]{@{}c@{}}0.8748\\ ±0.0003\end{tabular}} & \textbf{\begin{tabular}[c]{@{}c@{}}0.8705\\ ±0.0005\end{tabular}} & \begin{tabular}[c]{@{}c@{}}0.9005\\ ±0.0003\end{tabular} & \textbf{\begin{tabular}[c]{@{}c@{}}0.9208\\ ±0.0001\end{tabular}} & \textbf{\begin{tabular}[c]{@{}c@{}}0.9314\\ ±0.0002\end{tabular}} & \begin{tabular}[c]{@{}c@{}}0.6411\\ ±0.0027\end{tabular} & \textbf{\begin{tabular}[c]{@{}c@{}}0.6455\\ ±0.001\end{tabular}} & \textbf{\begin{tabular}[c]{@{}c@{}}0.6529\\ ±0.0011\end{tabular}} \\ \cline{2-14} 
 & FB & \begin{tabular}[c]{@{}c@{}}0.8258\\ ±0.0003\end{tabular} & \begin{tabular}[c]{@{}c@{}}0.8368\\ ±0.0006\end{tabular} & \begin{tabular}[c]{@{}c@{}}0.8427\\ ±0.0003\end{tabular} & \begin{tabular}[c]{@{}c@{}}0.8740\\ ±0.0004\end{tabular} & \begin{tabular}[c]{@{}c@{}}0.8725\\ ±0.0003\end{tabular} & \begin{tabular}[c]{@{}c@{}}0.8658\\ ±0.0003\end{tabular} & \begin{tabular}[c]{@{}c@{}}0.9019\\ ±0.0002\end{tabular} & \begin{tabular}[c]{@{}c@{}}0.9167\\ ±0.0004\end{tabular} & \begin{tabular}[c]{@{}c@{}}0.9284\\ ±0.0001\end{tabular} & \begin{tabular}[c]{@{}c@{}}0.6466\\ ±0.0021\end{tabular} & \begin{tabular}[c]{@{}c@{}}0.6385\\ ±0.0011\end{tabular} & \begin{tabular}[c]{@{}c@{}}0.6532\\ ±0.0011\end{tabular} \\ \hline
\end{tabular}
\end{table*}
}

\subsection{Recommendation Performance Comparison}
\label{sec: performance_comparison}

Table~\ref{tab:result1} presents the experimental results, showing the average RMSE and AUC. It is important to note that an AUC increase at the 0.001 level is considered statistically significant \cite{wide_deep,DeepFM}. Based on the data in the table, we can draw the following conclusions in response to \textbf{RQ1}

(1) In the context of experience replay, our findings align with existing study \cite{Gdumb} that a simple randomly replaying strategy (GDumb), can effectively mitigate catastrophic forgetting. However, our proposed method MEGG consistently outperforms all experience-replaying methods across all experiment settings, as evidenced by the average RMSE and AUC metrics. There are only several exceptions in that GDumb slightly surpasses MEGG.

(2) In the broader context of incremental recommendation, the MEGG method consistently outperforms the current state-of-the-art methods, including the knowledge distillation-based IncCTR method and the meta-learning-based SML method, across various experimental setups.

(3) When compared with Full-Batch, the MEGG method achieves nearly identical recommendation results in most experiments and even surpasses Full-Batch in some instances. This indicates that our sampling strategy can make data reservoir $D$ contain valuable information equivalent to the full batch data for model learning. This phenomenon shows that MEGG can ensure that the recommendation performance does not decline but also significantly reduces storage and computational overhead.

\subsection{Replaying Component for Performance Enhancement}
\label{sec: integration}
As mentioned in Section~\ref{sec: incremental_learning}, one of the advantages of experience replay-based methods is their potential for integration with other incremental learning methods may result in further performance enhancement. For example, in the case of IncCTR, there exists a Data Module which is designed for replaying. However, the author has yet to find a suitable strategy to implement it. Therefore, in this section, we investigate the improvement in recommendation performance by combining the IncCTR and SML methods with our MEGG approach. \textcolor{mygreen}{Specifically, we modify the training data used during each incremental stage of fine-tuning, replacing $D'$ with the data $D$ maintained by the Data Reservoir of our MEGG method, while keeping the other components of the IncCTR and SML methods unchanged.} Similarly, we also apply this combination with GDumb. We report the mean and standard deviation (to four decimal places, with the last two digits displayed) of the experimental results, as well as the improvement relative to the no-combination, in Table~\ref{tab:enhanced_performance}. According to the experimental results, we can make the following observations, which address \textbf{RQ2}.

(1) Performance Enhancement. For the WDL and DCN models, the recommendation performance of IncCTR and SML can be further improved by incorporating the replay method. However, for the NFM model, it is more suitable to employ a single incremental learning method exclusively.

(2) Comparison with GDumb. Across all experimental setups, the enhanced recommendation performance of IncCTR and SML methods achieved by combining the MEGG method surpasses that achieved by combining the GDumb method. This further underscores the superiority of the MEGG in migrating catastrophic forgetting.

\begin{table*}[]
\caption{Recommendation performance of combined methods and the performance enhancement compared with no-combination.}
\label{tab:enhanced_performance}
\centering
\vspace{-0.2cm}
\begin{tabular}{cc|ccc}
\hline
\multirow{2}{*}{Metric} & \multirow{2}{*}{Method} & \multicolumn{3}{c}{MovieLens-1M} \\ \cline{3-5} 
 &  & WDL & DCN & NFM \\ \hline
\multirow{4}{*}{\begin{tabular}[c]{@{}c@{}}AVG \\ RMSE\end{tabular}} & \begin{tabular}[c]{@{}c@{}}IncCTR\\ +MEGG\end{tabular} & \textbf{\begin{tabular}[c]{@{}c@{}}0.9435(09)\\ ↑1.3900\%\end{tabular}} & \textbf{\begin{tabular}[c]{@{}c@{}}0.9321(10)\\ ↑0.7454\%\end{tabular}} & \textbf{\begin{tabular}[c]{@{}c@{}}0.9559(26)\\ ↓1.8215\%\end{tabular}} \\
 & \begin{tabular}[c]{@{}c@{}}IncCTR\\ +GDumb\end{tabular} & \begin{tabular}[c]{@{}c@{}}0.9483(06)\\ ↑0.8884\%\end{tabular} & \begin{tabular}[c]{@{}c@{}}0.9346(06)\\ ↑0.4792\%\end{tabular} & \begin{tabular}[c]{@{}c@{}}1.0041(24)\\ ↓6.9557\%\end{tabular} \\ \cline{2-5} 
 & \begin{tabular}[c]{@{}c@{}}SML\\ +MEGG\end{tabular} & \textbf{\begin{tabular}[c]{@{}c@{}}0.9471(11)\\ ↑0.8065\%\end{tabular}} & \textbf{\begin{tabular}[c]{@{}c@{}}0.9281(03)\\ ↑1.2765\%\end{tabular}} & \textbf{\begin{tabular}[c]{@{}c@{}}0.9593(18)\\ ↓2.5551\%\end{tabular}} \\
 & \begin{tabular}[c]{@{}c@{}}SML\\ +GDumb\end{tabular} & \begin{tabular}[c]{@{}c@{}}0.9507(11)\\ ↑0.4294\%\end{tabular} & \begin{tabular}[c]{@{}c@{}}0.9331(03)\\ ↑0.7446\%\end{tabular} & \begin{tabular}[c]{@{}c@{}}0.9724(13)\\ ↓3.9555\%\end{tabular} \\ \hline
\multirow{4}{*}{\begin{tabular}[c]{@{}c@{}}AVG\\ AUC\end{tabular}} & \begin{tabular}[c]{@{}c@{}}IncCTR\\ +MEGG\end{tabular} & \textbf{\begin{tabular}[c]{@{}c@{}}0.8371(06)\\ ↑1.9486\%\end{tabular}} & \textbf{\begin{tabular}[c]{@{}c@{}}0.8457(08)\\ ↑1.5734\%\end{tabular}} & \textbf{\begin{tabular}[c]{@{}c@{}}0.8271(13)\\ ↓1.1355\%\end{tabular}} \\
 & \begin{tabular}[c]{@{}c@{}}IncCTR\\ +GDumb\end{tabular} & \begin{tabular}[c]{@{}c@{}}0.8306(03)\\ ↑1.1570\%\end{tabular} & \begin{tabular}[c]{@{}c@{}}0.8382(06)\\ ↑0.6726\%\end{tabular} & \begin{tabular}[c]{@{}c@{}}0.8117(10)\\ ↓2.9763\%\end{tabular} \\ \cline{2-5} 
 & \begin{tabular}[c]{@{}c@{}}SML\\ +MEGG\end{tabular} & \textbf{\begin{tabular}[c]{@{}c@{}}0.8331(07)\\ ↑1.5851\%\end{tabular}} & \textbf{\begin{tabular}[c]{@{}c@{}}0.8426(04)\\ ↑1.1646\%\end{tabular}} & \textbf{\begin{tabular}[c]{@{}c@{}}0.8314(16)\\ ↓1.0709\%\end{tabular}} \\
 & \begin{tabular}[c]{@{}c@{}}SML\\ +GDumb\end{tabular} & \begin{tabular}[c]{@{}c@{}}0.8283(05)\\ ↑0.9998\%\end{tabular} & \begin{tabular}[c]{@{}c@{}}0.8362(04)\\ ↑0.3962\%\end{tabular} & \begin{tabular}[c]{@{}c@{}}0.8207(07)\\ ↓2.3441\%\end{tabular} \\ \hline
\end{tabular}
\vspace{-7pt}
\end{table*}

\begin{figure*}[!tph] 
  \centering
    \includegraphics[width=5.1in]{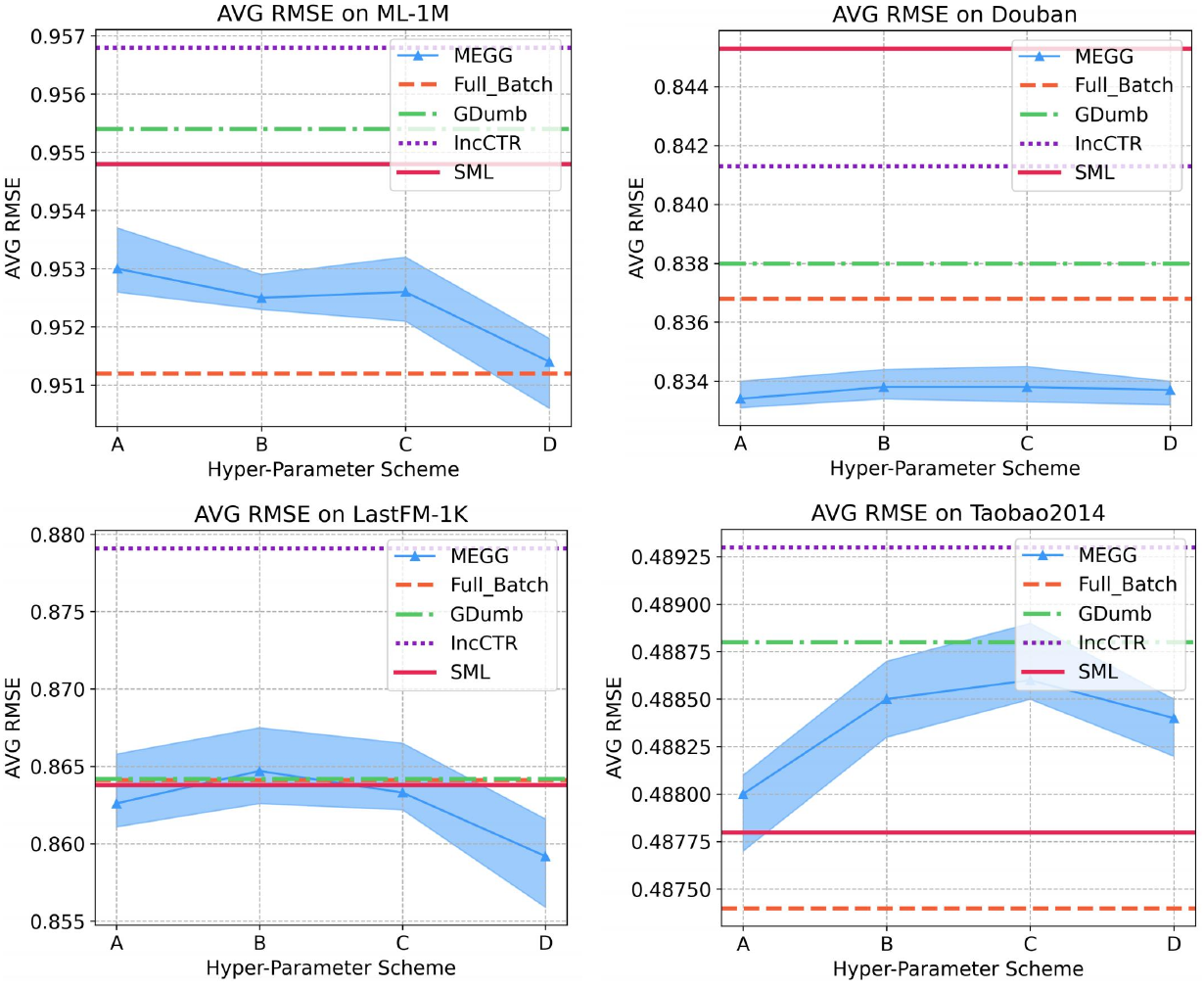}
    \vspace{-2mm}
   \caption{The average RMSE with different hyper-parameter schemes. Other baseline methods' performances are depicted as horizontal lines.}
    \label{fig:hyper_parameter}
\end{figure*}

\subsection{Efficiency Analysis}
\label{sec: efficiency_analysis}
\begin{figure}[t]
  \centering
    \includegraphics[width=3.0in]{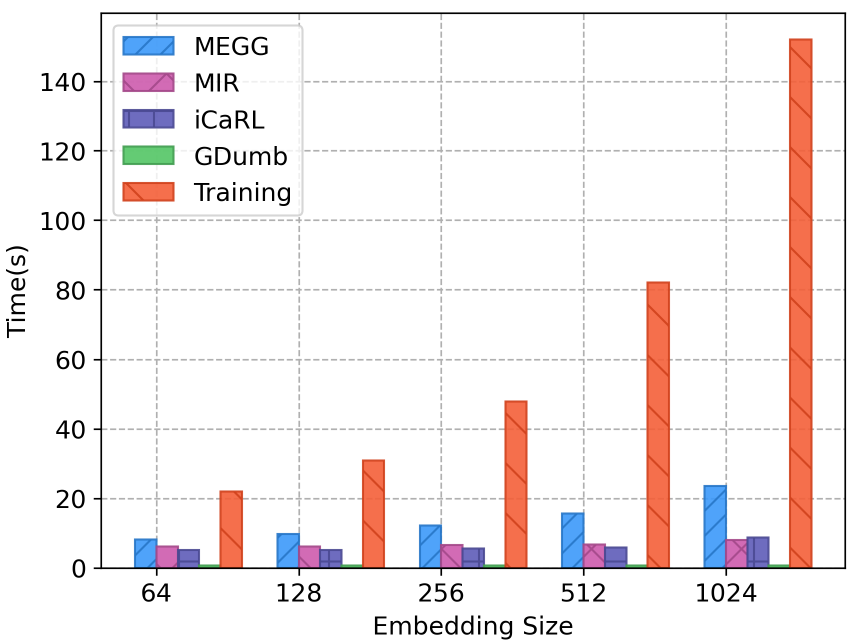}
    \vspace{-2mm}
   \caption{Time costed on model training and sampling of different methods.}
    \label{fig:time_cost}
\end{figure}

To answer \textbf{RQ3}, we progressively increase the embedding size of models and record the time cost for model training and sampling of different sampling methods. The experiment is conducted on the WDL model with the Taobao2014 dataset and the results are illustrated in Fig.~\ref{fig:time_cost}. It can be observed that as the embedding size increases, the model training time escalates a lot, and the sampling time consumed by MEGG is just slightly increased. We believe that such efficiency owing to the parameter selection method proposed in Section~\ref{sec:parameter_selection}. Even with an embedding size of 1024, the number of parameters involved in calculating GGscore remains relatively small. Admittedly, when juxtaposed with several other methods, the efficiency of MEGG at larger embedding sizes is somewhat inferior. However, in real-world industrial applications, the embedding size typically remains below 128. Given the improvements in prediction brought by the MEGG method (even minor improvements, such as 0.1\%, are substantial in recommendation systems as per \cite{huawei_IRS}), the additional computational overhead introduced by MEGG is deemed acceptable.

\subsection{Hyper-parameter Analysis}
\label{sec: hyperparameter_analysis}
In this section, we address \textbf{RQ4}. We represent the model parameters, trained after each epoch, as $\theta_{1},\theta_{2},\dotsb \theta_{5}$. The hyper-parameter configurations in Algorithm~\ref{alg:algorithm} are denoted as scheme A, where $\hat{\theta}^{'}=\theta_{4}$, $\hat{\theta}=\theta_{5}$, and $\mathbf{V}=\nabla_{\hat{\theta}}L(D,\hat{\theta})$. To facilitate a comprehensive comparison, we introduce an additional three schemes, namely B, C, and D. The specifics of these four schemes are delineated as follows:

A. $\hat{\theta}^{'}=\theta_{4}$, $\hat{\theta}=\theta_{5}$, $\mathbf{V}=\nabla_{\hat{\theta}}L(D,\hat{\theta})$. 

B. $\hat{\theta}^{'}=\theta_{3}$, $\hat{\theta}=\theta_{4}$, $\mathbf{V}=\nabla_{\hat{\theta}}L(D,\hat{\theta})$

C. $\hat{\theta}^{'}=\theta_{3}$, $\hat{\theta}=\theta_{5}$, $\mathbf{V}=\nabla_{\hat{\theta}}L(D,\hat{\theta})$

D. $\hat{\theta}^{'}=\theta_{4}$, $\hat{\theta}=\theta_{5}$, $\mathbf{V}=\nabla_{\hat{\theta}}L(D\cup D^{'},\hat{\theta})$

In the experimental setup employing the WDL model across four distinct datasets, we undertake a comparative analysis of recommendation performance amongst the aforementioned four schemes. The outcomes are graphically represented in Fig.~\ref{fig:hyper_parameter}. Notably, all four unique parameter configurations demonstrate outstanding predictive efficacy. Nevertheless, our findings also reveal that diverse parameter configurations exert a substantial influence on the MEGG method's performance, underscoring the necessity to judiciously select appropriate hyper-parameters in accordance with practical scenarios.

\begin{figure}[t]
  \centering
    \includegraphics[width=3.0in]{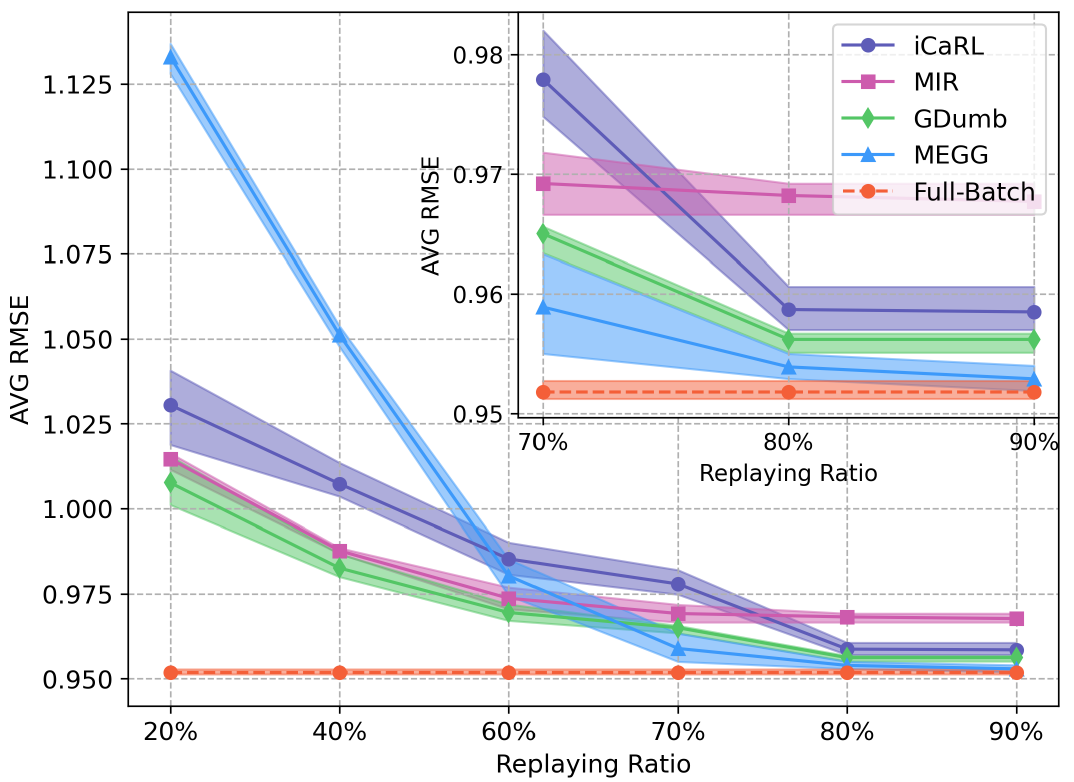}
    \vspace{-2mm}
   \caption{The average RMSE of different replay ratios.}
    \label{fig:replaying_ratio}
\end{figure}

\subsection{Replay Ratio Analysis}
\label{sec: sampling_size_analysis}

This section addresses \textbf{RQ5}. We deviate from the approach of selecting $M-M^{'}$ historical samples at each stage, and instead, employ different replay ratios including $20\%, 40\%, \ldots 90\%$ of $M$ to govern the size of replayed samples. Utilizing the WDL model and the ML-1M dataset, we evaluate the recommendation performance of MEGG in comparison with iCaRL, MIR, and GDumb, with the outcomes graphically represented in Fig.~\ref{fig:replaying_ratio}. Based on these results, we observe that when the replay ratio exceeds 70\%, MEGG outperforms other replaying methods. Furthermore, when the replay ratio surpasses 80\%, it exhibits performance comparable to the Full-Batch method.

\section{Conclusion}
In this study, we present a pioneering experience replay-based incremental learning approach called MEGG, specifically designed for deep neural network-based recommendation models. It effectively addresses the challenge of catastrophic forgetting of user's long-term preferences faced by online dynamic recommender systems, enabling recommendation models consistently maintain high recommendation performance. We empirically demonstrate the effectiveness, expandability, efficiency, and robustness of MEGG.

\bibliography{sn-bibliography}

\end{document}